\documentclass[12pt]{book}

\usepackage[dvips]{graphicx,color,epsfig}
\usepackage{makeidx,tsukuba,here}

\makeauthorindex
\makeindex

\begin{document}

\BookTitle{\itshape The 28th International Cosmic Ray Conference}
\CopyRight{\copyright 2003 by Universal Academy Press, Inc.}
\pagenumbering{arabic}

\chapter{Shower Studies at around $10^{18}$ eV with the Surface Detector
of the Pierre Auger Observatory}

\author{P.L.Ghia$^1$ for The Pierre Auger Collaboration$^2$\\
{\it (1) CNR-IFSI, Torino and INFN-LNGS, Assergi, Italy\\
\it (2) Obs. Pierre Auger, Av. San Martin Norte 304,
(5613), Malargue, Argentina}\\
}

\section*{Abstract}
Three months of data taken by 
the Engineering Array Surface Detector of the Pierre Auger
Observatory have been analyzed with the aim of setting a procedure for the
study of showers at $E_0\approx 10^{18}$ eV. We present the results, concerning:
(i) the selection procedure
and the check of its efficiency; (ii) the event rate and its
stability; (iii) the characteristics of such events (i.e., angular
direction, energy estimator $S(1000)$, rise time vs core
distance, correlation with pressure) showing their consistency
with the expectations. The performances of the whole Surface
Detector of the Auger Observatory are discussed.
\section{Introduction}
The study of cosmic rays around $10^{18}$ eV with the surface detector
of the Auger Observatory will be of main significance for:
a) the verification and extension of the CR anisotropy
measurement reported by AGASA [1], SUGAR [2] and Fly's Eye [3]
from the directions of the Galactic Center and Disk; b) the
measurement of the CR flux in an energy range interconnected with
other experiments; c) the possibility of improving the cross
calibrations with the Auger fluorescence detector, which naturally
operates at lower threshold.  The Auger southern site surface
detector, under construction in the province of Mendoza,
Argentina, will consist of 1600 water Cherenkov detectors (WCD),
on a triangular grid with a 1.5 km spacing, covering an area of
about 3000 km$^2$; a subset of it, the ``Engineering Array'' (EA),
including 
32 completely instrumented stations,
has been in operation since 2002. We
present here the analysis of three months of data taken between
May and July 2002, to set a procedure for the selection
of showers around $10^{18}$ eV.
\section{Analysis and results}
To define the most appropriate trigger configuration to
select events with $E_0\approx 10^{18}$ eV, we conducted a
preliminary study, using the data from a WCD prototype operated
in correlation with AGASA ($E_0>10^{18}$ eV and $\theta<45^\circ$)
[4]. The analysis [5] consisted of the simulation of the Auger
trigger and its application to the WCD
data after proper conversion to the Auger scale. This allowed the
derivation of
the single tank trigger efficiency as a function of its
rate, $f_t$, core distance, $r$, and energy, $E_0$.
The array trigger efficiency, $\eta$, obtained from such data, 
is shown in Tab.1 for different $f_t$ and $E_0$.
\begin{table}[H]
\vfill \begin{minipage}{.47\linewidth}
\begin{center}
\vspace{-0.3cm} \mbox {
\begin{tabular}{cccc}
\hline
$f_t$[Hz] & $E_0$[eV]  & $\eta_{3-fold}$ & $\eta_{4-fold}$    \\
\hline
20 & 1-2$\cdot 10^{18}$  & 0.25     & 0.01 \\[0.1cm]
20 & 2-4$\cdot 10^{18}$  & 0.45     & 0.07 \\[0.1cm]
2 & 1-2$\cdot 10^{18}$  & 0.23     & 0.01 \\[0.1cm]
2 & 2-4$\cdot 10^{18}$  & 0.40     & 0.05 \\[0.1cm]
\hline
\end{tabular}
}
\end{center}
\vspace{-1pc}
\caption{\it Array trigger efficiencies for different $E_0$ and
$f_t$.}
\end{minipage}\hfill
\begin{minipage}{.50\linewidth}
\begin{center}
\vspace{-0.3cm} \mbox {
\begin{tabular}{cccc}
\hline
$f_t$[Hz] & $\Delta t$[$\mu$s] & $N_{ch}$(/yr) & $N_{exp}$(/yr)\\
\hline
20 & 25 & $1.5\cdot 10^6$ & $3\cdot 10^4$ \\[0.1cm]
20 & 5 & $6\cdot 10^4$ & $3\cdot 10^4$ \\[0.1cm]
2 & 25 & $1.5\cdot 10^3$ & $2.8\cdot 10^4$ \\[0.1cm]
2 & 5 & $60$ & $2.8\cdot 10^4$ \\[0.1cm]
\hline
\end{tabular}
}
\end{center}
\vspace{-1pc}
\caption{\it Full Auger: expected events, $N_{exp}$, and accidental 3-fold
ones, $N_{ch}$, for different $f_t$ and $\Delta t$.}
\end{minipage}\vfill
\end{table}
Due to the low efficiency of 4-fold mode, 3-fold events have to be
used. The background rate due to 3-fold random coincidences
for the whole Auger detector is given in Table 2 together with the rate of
expected CR events, $N_{exp}$, with $10^{18}<E_0<2\cdot 10^{18}$
eV, evaluated using the CR spectrum [6],
the Auger acceptance ($A\Omega \approx 4700$ sr km$^2$,
$\theta<45^\circ$) and $\eta$.

The extraction algorithm is firstly based on the selection of 3
nearby triggered tanks in a equilateral pattern. For the Auger-EA ($f_t
\approx 20$ Hz and coincidence window $\Delta t \approx 25 \mu$s)
the measured rate, $N_{sel} \approx 30$/day, is consistent with
that expected from chance coincidences. 
\begin{figure}[H]
\vfill \begin{minipage}{.47\linewidth}
\vspace{-0.6cm}
To reduce the
accidental rate, we lower (through software cuts) both $\Delta t$
(to $5\mu$s) and $f_t$ (to 2 Hz), so that $N_{ch}\approx 1.5\cdot 10^{-3} $/day.

With the described selection criteria, 
in the period from May 1st through July 11th 2002,
we extracted 487 events (416 with $\theta<45^\circ$). Fig.1 (top)
shows the daily rate ($\theta<45^\circ$), for days of operation
with duty cycle greater than 98\%. The daily rate distribution is
shown in fig.1 (bottom): the average is $6.3\pm0.4$/day, the
distribution being compatible ($\chi^2=1.5/d.f.$) with a poissonian
one. 
\end{minipage}\hfill
\begin{minipage}{.50\linewidth}
\vspace{-1.cm}
\begin{center}
\mbox{\epsfig{file=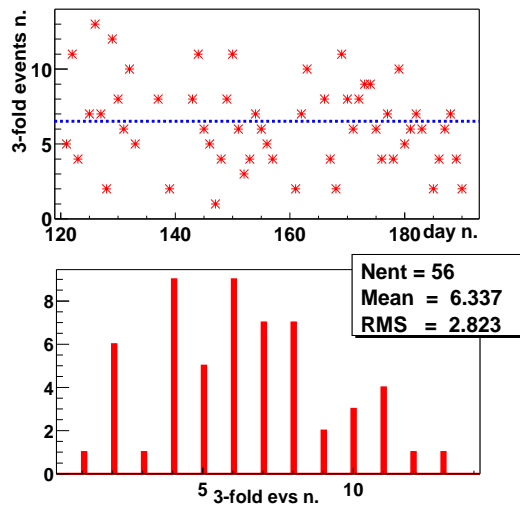,height=7.cm,width=7.2cm}}
\end{center}
\vspace{-1.5pc} \caption{\it 3-fold event daily rate
($\theta<45^\circ$) vs day number (top) and its distribution
(bottom).}
\end{minipage}\vfill
\end{figure}
The measured rate is then compared with the expected one due to CRs.
Taking into account the contributions from different energies
($2.5\cdot 10^{17}<E_0<4\cdot 10^{18}$ eV),
we evaluate it using spectrum [6],
including the Auger-EA acceptance ($A\Omega
\approx 51$ sr km$^2$, $\theta<45^\circ$) and the array trigger
efficiency $\eta$. 
$\eta$ is obtained
using the WCD prototype data (converted to the Auger-EA
scale), in correlation with the AGASA EAS, for the different energy
intervals, with an extrapolation to $E_0<10^{18}$ eV.
The total predicted rate amounts to $N_{exp}=5\pm 1$/day, consistent
with the measured one, within the
systematic uncertainties (we remind that the results are based on trigger
efficiencies evaluated at the AGASA energy scale).

\begin{figure}[H]
\vspace{-0.8cm}
\begin{center}
\epsfig{file=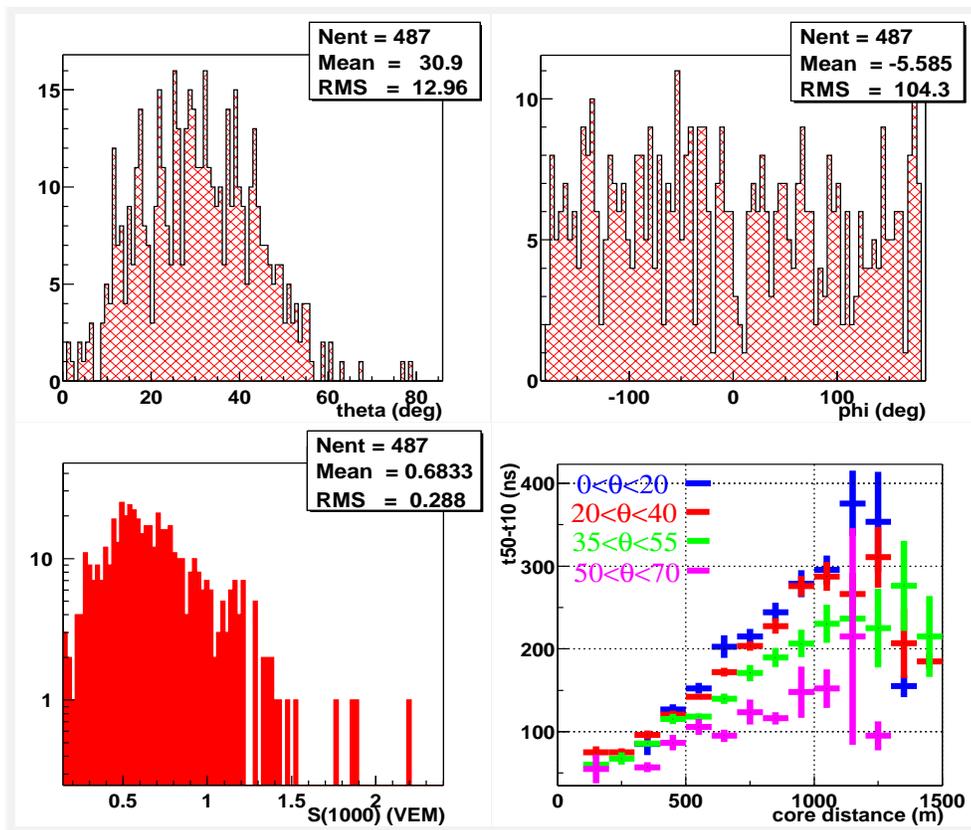,height=11cm,width=13cm}
\end{center}
\vspace{-1.5pc}
\caption{Distributions of reconstructed shower parameters for the 487
selected events: zenithal angle (top left), azimuthal angle (top right),
$S(1000)$ (bottom left) and rise time vs core distance (bottom right)
for four different intervals of $\theta$.}
\end{figure}
The extracted events have been reconstructed, following the
prescriptions given in [7].
The distributions of the arrival angles $\theta$ and
$\phi$ are given in fig. 2 (top left and right, respectively), 
showing good consistency
with the expectations. The risetimes from 10\% to 50\% levels of
the signal ($t_{50}-t_{10}$) vs core distance ($r$) in four
intervals of $\theta$ are shown in the same figure (bottom right), 
being in good agreement
with simulated data [8] and real ones from Haverah Park [9]. 
Concerning primary energy, an
estimator, $S(1000)$ (i.e., the signal in Vertical Equivalent Muons at
$r=1000$m) corrected for $\theta$, is shown in fig. 2 (bottom
left). The conversion from $S(1000)$ to primary energy
is subject of intense study by the Auger Collaboration (see e.g. [7]):
currently, the peak
of the $S(1000)$ distribution corresponds to primary energies
ranging from 0.4 to 0.9$\cdot 10^{18}$ eV (consistent with the present knowledge
of the primary intensity at such energies, when compared
with the measured rates).
Lastly, the behavior of the event rate vs atmospheric pressure,
$P$, has been studied. The pressure coefficient obtained is $k=1/N
(dN/dP)=-0.011\pm 0.005$ mbar$^{-1}$ (significant at
2$\sigma$), in agreement with Haverah Park [10].

\section{Conclusions}
Three months of data taken by the Auger-SD EA in a stable
configuration have been analyzed, with the following outcome:\\
- an algorithm for the extraction of events around $10^{18}$ eV,
based on geometry and single tank trigger condition corresponding
to $f_t\approx$2 Hz (to reduce the random coincidences to the
percent level of the physical event rate) has been tuned and
tested using a WCD prototype operated in connection
with AGASA;\\
- the event rate is consistent with the expectations
from CR showers due to primaries of energy around $10^{18}$ eV;\\
- the shower parameters (distributions of zenith and azimuthal
angles, $S(1000)$, rising time vs core distance
and vs zenith angle, correlation with pressure) are consistent
with the expected ones;\\
- assuming for the full Auger-SD $\eta=0.23$, and using the
spectrum [6], we get $N_{exp}(10^{18} {\textrm{eV}}<E_0<2\cdot
10^{18} {\textrm{eV}})\approx 2.8\cdot 10^4$ ev/yr (in the 
AGASA energy scale). The detection level of the AGASA anisotropy
($A\approx4$\%), assuming poissonian fluctuations, would thus be
$\approx$ 4.5 s.d./yr (in harmonic analysis). Concerning stability,
representing a further requirement for the measurement,
within the present statistics, non poissonian daily instability
effects can be verified to be $<$ 6-7\%.

\section{References}
\re
1.\ Hayashida N. et al.\ 1999, Astropart. Phys. 10, 303
\re
2. Bellido J.A. et al.\ 2001, Astropart. Phys. 15, 167
\re
3. Bird D.J. et al.\ 1999, Ap. J. 511, 739
\re
4.\ Sakaki N. et al.\ 1997, Proc. 25th ICRC 5, 217
\re
5.\ Ghia P.L. and Navarra G.\ 2003, Auger-GAP note 2003-007
\re
6.\ Nagano M. and Watson A.A.\ 2000, Rev. Mod. Phys. 72, 689
\re
7.\ Billoir P.\ 2002, Auger-GAP note 2002-044, 2002-075
\re
8.\ Sciutto S.\ 2002, private communication
\re
9.\ Watson A.A. and Wilson J.G.\ 1974, J.Phys.A 7, 1199, Walker R. and
Watson A.A.\ 1981, J.Phys.G 7, 1297 
\re
10.\ $k=-0.007\pm 0.001$ mbar$^{-1}$, Edge D.M. et al.\ 1978, J.Phys.G 4, 133

\endofpaper
\end{document}